\documentclass[conference]{IEEEtran}
\ifCLASSINFOpdf
  \usepackage[pdftex]{graphicx}
\else
\fi
\usepackage[cmex10]{amsmath}
\begin{document}
%
\title{Flying Capacitor Cell Equalization for Li-ion Automotive Battery Stacks}

\author{\IEEEauthorblockN{Manish Ramaswamy}
\IEEEauthorblockA{Electronics \& Control Systems\\
American Axle \& Manufacturing\\
Detroit, MI 48211\\
Email: manny.swamy@ieee.org}
}


%


\maketitle

\begin{abstract}
 The automotive industry is fast evolving to Li-ion chemistries, which have more favorable power, energy density, and efficiency. To meet the demands of greater electric ranges, parallel strings of batteries are required to increase the overall system capacity.  Differences in chemical characteristics, internal resistance, and operating conditions can cause variations in remaining cell capacity, decreasing the total battery storage capacity over time, shortening the battery lifetime and eventually damaging the cells. Cell equalization tries to restore all the cells in the pack to an equal state of charge in order to prolong the battery lifetime and to ensure safe battery operations. This work presents an active charge equalization scheme with  a flying capacitor to shuttle charge between the unbalanced cells in a parallel battery pack.  The theoretical framework is accompanied by MATLAB simulations on a twelve cell pack in series/parallel configuration supporting the validity of the chosen approach.
\end{abstract}


%
\IEEEpeerreviewmaketitle

\section{Introduction}
A major issue in rechargeable battery research is efficient charging and discharging because it leads to short charging time and extends the life span of the battery pack. Lithium-ion batteries need special protection from overcharge and overdischarge, because if the single cell voltage becomes higher than 4.5 V,  it would cause a breakdown of the active materials within the cells or in the worst case cell explosion, while if the voltage goes under 2.3 V, internal chemical reactions cause the cell to irreversibly lose large part of its capacity [1].

To meet the power demand from a P/H/EV, batteries are constructed into a pack. First, multiple cells are connected in series (form a string) to provide higher voltage. The strings can then be connected in parallel to provide higher current capacity. In this configuration, factors such as different internal resistance, varied capacity, temperature gradients, and general differences caused by manufacturing process tolerance can lead to imbalance in terms of State of Charge (SoC) and cell voltages. 

To enhance the electrochemical uniformity of the battery cells for increased  battery lifetime and to insure safe battery operations, individual cells need to be maintained at an equalized  charge level. Therefore, equalization capability is an important design consideration for applications of Li-ion battery in powering electric vehicles. Several equalization methods have been proposed in literature [2]. A simple and most widely used technique for cell balancing is charge shunting [3] and [4], as it is easy to implement. This method uses a resistor to shunt the charging current whenever there is any excess charge inside a cell, and  the SoC is estimated by measuring the terminal voltage. Cell balancing can be done using  capacitors, and several techniques are found in [4], [5],  and [6]. The basic idea is to use a capacitor to remove and hold the additional charge when a cell tends to be overcharged.  The capacitor also transfers the charge to another partially charged cell. A variation is to use a``flying capacitor." The cell with the highest SoC will charge the flying capacitor, and the cell with the lowest SoC will receive that charge. Some control schemes are necessary to select the highest and lowest charged cells.

In addition to using flying capacitors, other active cell balancing methods include discharge equalizing system, like multi-output transformers [7], charge equalizing systems, like distributed Cuk converter [8], [9], and bidirectional equalizing systems, like switched capacitor or inductor circuit [10]. Each one of those schemes has its advantages and drawbacks, in terms of equalization speed, circuit complexity, number of parts needed and rating of the part to be used, particularly the current rating of the switching components and the voltage rating of the diodes.

From the above discussion, we see that all  of the active cell balancing methods use some sort of charge routing or  storage device, say a capacitor, inductor, or a combination of both  to shuttle energy between cells. A majority of what has appeared in the literature is for series strings of batteries. In this research, our focus will be to develop a scheme to shuttle charge on parallel battery packs which finds application in EVs and PHEVs.

\section{Battery Pack Model}

In most current HEV architectures, the battery pack is made up of a single string of cells. Modeling and simulating a string of cells is a relatively easy task. PHEV battery architectures vary slightly from HEV battery architectures. PHEV battery packs are most commonly configured as parallel strings in order to increase overall pack capacity  to support the necessary range requirements for the vehicle. It is for this reason that we focus on battery pack in a parallel configuration. The equivalent circuit model used in this work has been derived in [11]. One of the major concerns when designing a parallel pack is the current split which is the current input as seen by each string in the pack. If the cells were identical, the computation is simply a matter of dividing the overall input current by the number of batteries. However, since each cell is a dynamic system and due the fact that the laws of physics must be upheld, we must derive representation that allows for current splits between the strings to be calculated during the runtime of the simulation. The generalized form is derived to be

\begin{equation}
\left[
 \begin{matrix}
  \alpha_1 \\
  \alpha_2 \\
  \alpha_3\\
  \vdots \\
  \vdots\\
   \alpha_n
  
  \end{matrix}
\right]  = \left[
 \begin{matrix}
  \Phi_1 & -\Phi_2 &0 &\cdots& \cdots & 0  \\
  0 & \Phi_2 & -\Phi_3 & \cdots & \cdots & 0\\
  0 & 0 & \Phi_3 & -\Phi_4 & \cdots & 0 \\
  \vdots & \vdots & \vdots & \ddots & \ddots & \vdots \\
  \vdots & \vdots & \vdots &  \vdots & \phi_{n-1} & -\Phi_n \\
  1 & 1 & \cdots & \cdots & 1 & 1

  \end{matrix}
\right]^{-1} \left[
 \begin{matrix}
  \delta_1 \\
  \delta_2 \\
  \vdots\\
  \vdots \\
  \delta_n\\
   I
  
  \end{matrix}
\right]
\end{equation}
where $\delta_1 = \Gamma_2 - \Gamma_1 $, $\delta_2 = \Gamma_3 - \Gamma_2 $, $\delta_3 = \Gamma_n - \Gamma_{n-1} $, $\Phi \in \textbf{R}^{n \times n} $, $\Gamma \in \textbf{R}^{n \times 1} $, $\alpha_i$ is the current through each string, $n$ is the total number of parallel strings in the pack, and $I$ is the total current seen by the pack during simulation. $\Phi$ and $\Gamma$ matrices are defined as follows:

\begin{equation}
	\Phi_i = \displaystyle\sum_{i=1}^m-R_{0[i,j]} + R_{1[i,j]} (1-e^{- \beta_{1[i,j]^t}})+ R_{2[i,j]} (1-e^{- \beta_{2[i,j]^t}})
\end{equation}
\begin{equation}
	\Gamma_i = \displaystyle\sum_{j=1}^mE_{i,j} + \bar{V}_{c1[i,j]} (1-e^{- \beta_{1[i,j]^t}})+ \bar{V}_{c2[i,j]} (1-e^{- \beta_{2[i,j]^t}})
\end{equation}
where $m$ is the number of batteries in the $i^{th}$ string and $\bar{V}_{ck[i,j]}$ is is the voltage of the capacitor
containing the current  history of the battery. Also, through a simple application of Kirchhoff's current law, the forcing function can be computed as
\begin{equation}
	I = \alpha_1 + \alpha_2 + \cdots + \alpha_{n-1} + \alpha_n
\end{equation}

The battery terminal voltage drops significantly as the SoC approaches 0$\% $ and increases greatly as SoC approaches 100$\% $. In the middle portion, the relationship is approximately linear. Therefore, the OCV (Open Circuit Voltage) is modeled as a double exponential function given as:

\begin{equation}
V_{oc} = V_0 + \alpha (1-exp(-\beta z))+\gamma z+\zeta(1-exp(-\frac{\epsilon}{1-z}))
\end{equation}
where $z$ is the SoC in the defined range $z \in (0,1)$ and, $\alpha$, $\beta$, ..., $\epsilon$ are tunable parameters which can be found through optimization techniques.

The SoC of the battery is computed using a sumple Coulomb counting (current integration) as

\begin{equation}
\dot{z} = -\frac{1}{3600C_n}I(t)
\end{equation}
where $C_n$ is the battery capacity. Refer to [11] for more details on the model derivation.

The charge profile we use for this study is an extended Toyota Prius HEV current profile. This current profile is representative of a typical urban drive cycle  as would be seen by a commuter in a daily trip to or from the workplace. The active drive cycle is then followed by a 12 hour rest period presumably when the vehicle is not in service during work or overnight hours in which the battery cells can achieve a steady-state value and may or may not be subject to a cell balancing control system. 
\begin{figure}[htp]
  
  \centering
    \includegraphics[width=2in]{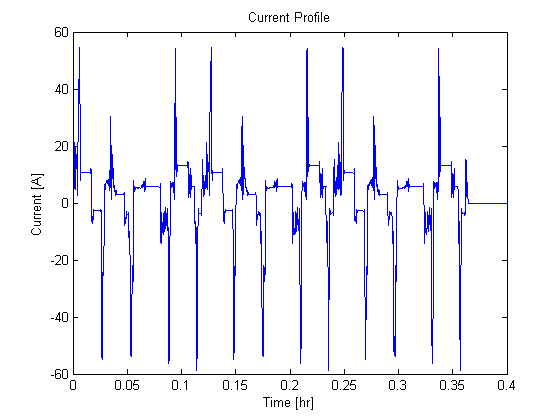}
\caption{Current profile.}
\end{figure}

\section{Active Cell Equalization}

One of the many methods to achieve cell balancing is with the use of a switched capacitor balancing circuit where a single capacitor is used to shift charges among batteries in a string. This ``flying capacitor" shown in Fig. 2, moves along the series string by intelligently closing switches around the desired cell to achieve balancing. Though intelligent control is needed in the case of flying capacitor to select which batteries are to be balanced, there is a significant reduction in the amount of electrical components thus reducing the costs. 

\begin{figure}[htp]
\centering
\includegraphics[width=2.5in]{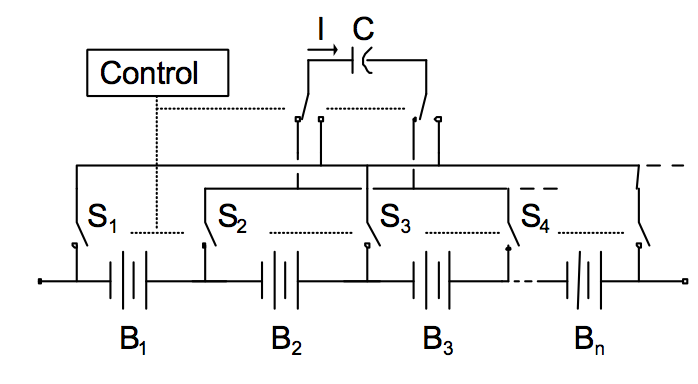}
\caption{Cell equalization with flying capacitor [4].}
\end{figure}

To increase the battery pack capacity, in this study we require strings of cells in parallel. The approach used to balance the cells in a parallel pack is an extension of the previous work by having the capacitor ``float" among the multiple strings and thereby eliminating the need to have a capacitor for each string.  Thus, we propose to use a single capacitor for the complete pack which has both series and parallel strings; the complete design is shown in Fig. 3. This work focuses on the feasibility and development of this active cell balancing system on a module. Even though temperatures greatly effect the performance and capacity of Li-ion cells, the implementation will not take into account any temperature changes. The pack is assumed to be used at an isothermal condition, an optimal temperature of 25C. However, the safety functions in a real EV should monitor the battery pack temperature and take appropriate actions  if it reaches a critical level.

\begin{figure}[htp]
\centering
\includegraphics[width=2.5in]{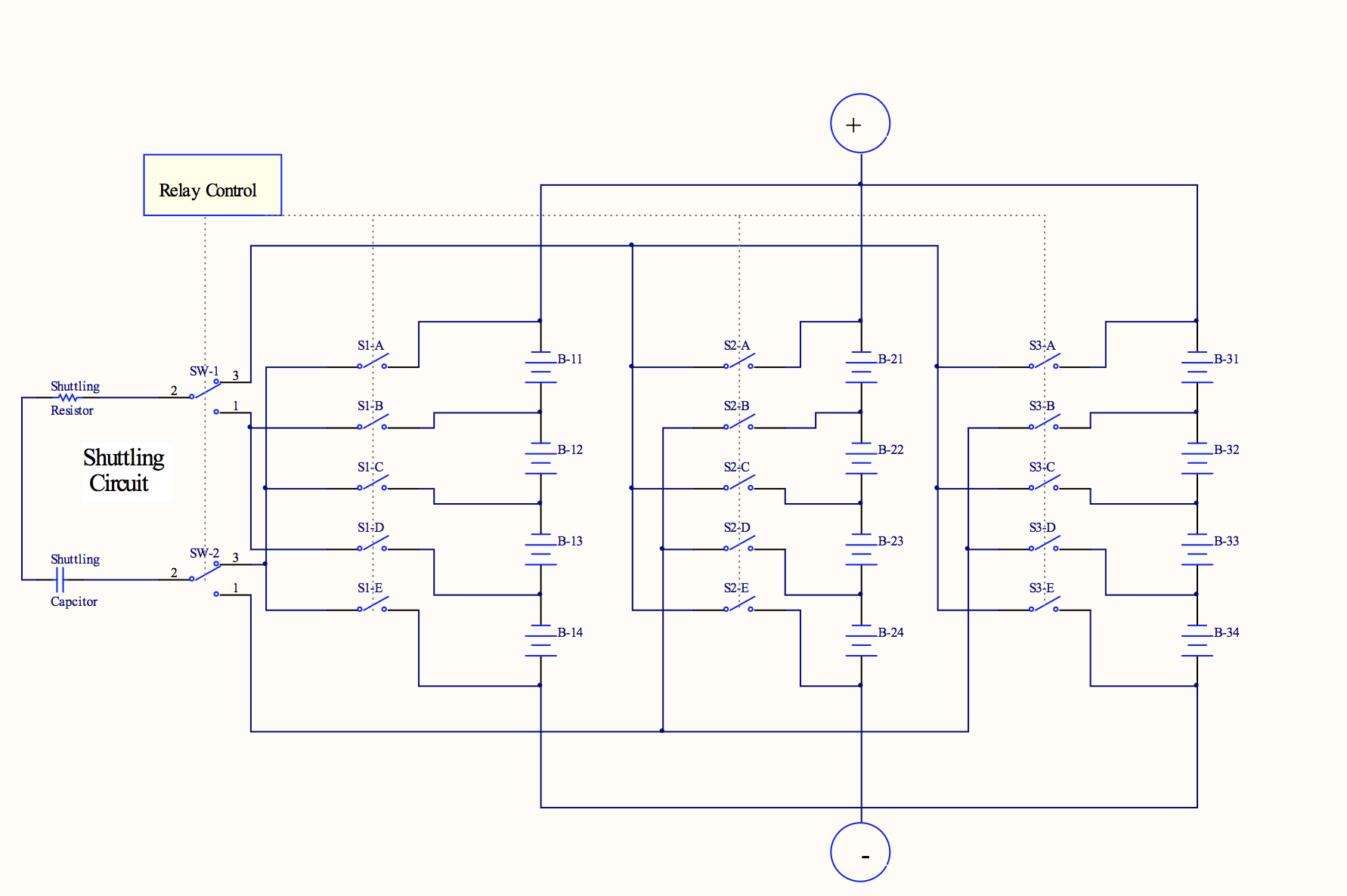}
\caption{Cell equalization control circuitry.}
\end{figure}

\subsection{Active Cell Balancing Control Structure}
The cell balancing circuitry for this study is comprised of a resistor $R$, and a capacitor $C$. The circuit connects to the battery selected in series as shown in Fig. 4. The voltage difference induces a current to flow in the circuit. The direction of the current will depend on  whether or not the voltage of the capacitor $V_c$ is higher or lower than the battery voltage $V_b$. The resistor $R$ is used to regulate the current between the capacitor and the battery.  

\begin{figure}[htp]
\centering
\includegraphics[width=1.3in]{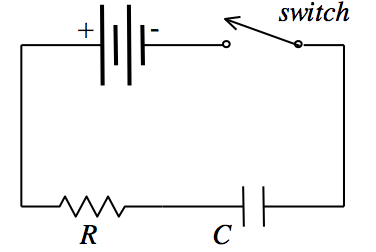}
\caption{Switched capacitor circuit. }
\end{figure}

Applying KVL to the circuit in Fig .4, we have
\begin{equation}
V_b(t) = Ri_c(t)+V_c(t)
\end{equation}
where $V_b(t) $ is the battery voltage at time $t$, $V_c(t)$ is the capacitor voltage at time $t$, and $i_c(t)$ is the current induced in the circuit.  The current can be expressed as either positive or negative, which indicates whether the battery is discharging or charging, respectively. The final value of the capacitor voltage after the current source has stopped charging the capacitor depends on the initial value of capacitor voltage and history of capacitor current. Solving  for capacitor voltage, we get
\begin{equation}
V_c(t) =\frac{1}{C}\int_{t_0^+}^t i_c(t) dt + V_c(t_0^-)
\end{equation}
where the integration of the current is over the present time step and $V_c(t_0^-)$ is the
voltage of the capacitor at the end of the previous time step. Substituting (8) into (7) yields
\begin{equation}
V_b(t) = Ri_c(t)+\frac{1}{C}\int_{t_0^+}^t i_c(t) dt + V_c(t_0^-)
\end{equation}

An expression for the current must be obtained before the capacitor voltage can be calculated and this computes as:

\begin{equation}
i_c(t) = \frac{(V_b(t) - V_c(t_{0-}))}{R} e^{\frac{-t}{RC}}
\end{equation}
The analytical solution for the voltage across the capacitor can now be solved for by substituting the induced current solution into
\begin{equation}
V_c(t) = \frac{1}{C} \int_{t_{0^+}}^t i_c(\tau)d\tau + V_c(t_{0-})C
\end{equation}
where the current is integrated across the present time step and $V_c(t_{0-})$ is the initial voltage that was solved for at the end of the previous time step.

From Kirchhoff's laws for parallel strings, we know the string voltages are equal. The batteries within the strings have slightly different internal characteristics, most notably internal resistances. This results in induced mesh currents between the parallel strings, which the battery pack model incorporates and solves for when determining the current split in (4). These mesh currents also occur during operation as well as when the battery pack is under no-load conditions. The induced current in the cell equalization circuit is also treated as a mesh current for the particular battery cell to which it is connected at that specific time interval. Therefore, this current must also be accounted for when calculating the voltage of the battery that is connected to the cell equalization circuit.

The current in the cell balancing circuit, $i_c(t)$, was found in (10). Therefore, when determining the voltage of the battery connected to the balancing circuit, the induced current from the balancing circuit and the mesh current, $i_m$, induced from the voltage imbalance of the parallel battery strings are additive as shown in Fig .5. Thus the observed battery current, $i_b(t)$, given as
\begin{equation}
i_b(t)  = i_c(t) + i_m(t)
\end{equation}

\begin{figure}[htp]
  
  \centering
    \includegraphics[width=1.75in]{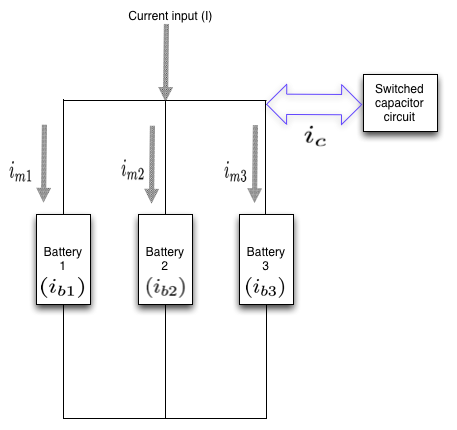}
\caption{Observed battery current.}
\end{figure}

\subsection{Active Cell Equalization Control Strategy}
Our main objective is to effectively balance the entire pack by making use of the active cell equalization circuit to transfer charge from a battery at a higher voltage to a battery at a lower voltage. This is accomplished by measuring the voltages of all the cells to determine which have a higher SoC and which have a lower SoC. A rule based control strategy is used which intelligently selects the highest charged cell and the lowest charged cell to balance, regardless of the location in the battery pack. The control algorithm is as follows: 

\begin{enumerate}
\item Check for no-load input
\item Set switching frequency to selected $RC$ time constant (discussed later).
\item Identify the lowest charged cell while charging the capacitor  from a cell with the maximum voltage.
\item Discharge the capacitor into the lower charged cell using the selected time constant. 
\item Repeat steps 3 and 4 until  SoC deviation of the selected battery cells is less than 2$\%$
\end{enumerate}

The simulation of the battery pack algorithm with the cell balancing circuit is carried out as follows:

\begin{enumerate}
\item Define battery characteristics for each individual cell.
\item Set up initial conditions for each battery (i.e. set SoC/OCV equal to represent a pack at equilibrium and capacitor voltages $(V_{ck[i,j]})$ set to 0V to simulate a completely relaxed battery).
\item Evaluate first simulation time step using first value of input current profile with initial battery parameters and SoC.
\item Begin simulating current profile.
\item Calculate battery parameters and OCV as a function of instantaneous battery SoC, current input, and individual battery characteristics.
\item Construct $\Phi$ and $\Gamma$ matrices using perturbed or base battery parameters.
\item Use the $\Phi$ and $\Gamma$ matrices to solve for the current split of the next simulation time step.
\item Locate the battery cells in the pack with the minimum and maximum voltages for balancing.
\item Evalute the balancing circuit capacitor voltage $V_c$ and the current $i_c$.
\item For each battery that is connected to the balancing circuit, evaluate the SoC, $V_{c1[i,j]}$, and $V_{c2[i,j]}$ for each battery in the pack.
\end{enumerate}

\subsection{Effect of Capacitor Size}

A main component of the charge balancing circuit is the super capacitor. Not only the cost but the value of the capacitor is a major concern as it has an effect on the settling time of the charge balancing process. To study the effect of capacitor size on the settling time, we run multiple simulations while keeping both the resistor size and switching time factor constant.
\begin{table}[htp]
\centering
\caption{$3P \times 4S$ End of Charge Balancing Time for $2\%$ SoC}
\begin{tabular}{|c|c|c|c|}

  \hline
   Capacitor (F) & Resistor ($\Omega$) & Switching time factor ($\delta)$ & Settling time (Hours)\\
   \hline
   20 & 0.05 & 0.5 & 11.7\\
    30 & 0.05 & 0.5 & 10.9\\
     50 & 0.05 & 0.5 & 10.4\\
    100 & 0.05 & 0.5 & 10.7\\
      150 & 0.05 & 0.5 & 11.0\\

   \hline
   
\end{tabular}

\end{table}

 Table 1 shows the simulation on a  $3P4S$ pack (string). The SoC imbalance at the end of the drive cycle was about 9.5$\%$ percent between the highest and lowest battery voltages, which is a higher percentage deviation than what we would see on typical urban  drive cycle and on a relatively new battery pack.  The simulations are run for final SoC deviation of 2$\%$. For this exercise, the resistor value and the switching time factor is held constant at 50m$\Omega$ and 0.5. Notice that when the capacitor size is increased, we see a decrease in the settling until 50F. However, increasing the capacitor size beyond this results in an increase in the settling time. This is due to an increase in the switching time of the charge balancing circuit.

\subsection{Effect of Resistor Variation}
Next, we analyze the contribution of the resistor component of the circuit. We choose three resistor values of 50m$\Omega$, 100m$\Omega$ and 200m$\Omega$. The capacitor size is fixed at 50F and the switching time factor is held constant at 0.5. The simulations are run on a full pack until the difference in the SoC is less than 2$\%$. The results are tabulated in Table 2. 

\begin{table}[htp]
\centering
\caption{$3P \times 4S$ End of Charge Balancing Time for $2\%$ SoC}
\begin{tabular}{|c|c|c|c|}

  \hline
   Capacitor  (F) & Resistor  ($\Omega$) & Switching time factor ($\delta)$ & Settling time (Hours)\\
   \hline
   50 & 0.05 & 0.5 & 10.4\\
 50 & 0.1 & 0.5 & 17.4\\
 50 & 0.2 & 0.5 & 26.2\\

   \hline
   
\end{tabular}

\end{table}
 Notice that we get the best balancing time using the least value of the resistor.  The maximum currents induced in the circuit for the $0.05 m \Omega $, $0.1 m \Omega $, $0.2 m \Omega $ resistors were $\pm$0.1, $\pm$0.05 and $\pm$0.04 amps, respectively. 
\subsection{Effect of Switching Time}
We now hold the resistor value and the capacitor size constant while varying the switching time. Simulations are run on a full pack until the difference in the SoC is less than 2$\%$. 
\begin{table}[htp]
\centering
\caption{$3P \times 4S$ End of Charge Balancing Time for $2\%$ SoC}
\begin{tabular}{|c|c|c|c|}

  \hline
   Capacitor (F) & Resistor ($\Omega$) & Switching time factor ($\delta)$ & Settling time (Hours)\\
   \hline
   50 & 0.05 & 0.5 & 10.4\\
 50 & 0.05 & 1 & 11.04\\
 50 & 0.05& 2 & 12.27\\

   \hline
   
\end{tabular}

\end{table}

Notice from Table 3 that the quickest balancing time is achieved for the most rapid switching time, as expected. 

\subsection{Energy Efficiency}

Another metric that helps in determining how well the system shuttles charge between the interconnected batteries is the energy efficiency.  This is an important metric because if a large amount of charge is lost for every unit transferred, it may not be worth shuttling the charge at all.  We calculate this metric by summing $(\sum)$ energy-related values over time in the manner
\begin{equation}
	\epsilon = \frac{\sum(\gamma)-\sum(\delta)}{\sum(\gamma)}
\end{equation}
where $\epsilon$ is the energy efficiency, $\gamma$ is the energy transferred and $\delta$ is the energy loss in passive elements.

To study the energy efficiency, we simulate two batteries connected in series with an initial SoC imbalance of $9.5\%$; one of the batteries is set at $60\%$ SoC and the other at $69.5\%$ SoC. Holding the initial battery parameters constant, numerous simulations were performed by varying the capacitor size, resistor value and the switching time. The initial capacitor value was held at 3.3 V. The simulation was run for a window of one hour with the capacitor sizes ranging from 30F to 180F and the resistor values ranging from 50m$\Omega$ to 500m$\Omega$. The switching frequencies were varied from $0.5 \tau$ to $3 \tau$ where $\tau$ is the time constant of the charge shuttling circuit. 

From Fig. 6 and 7, we see that the least resistor value and the switching frequency of $0.5 \tau$ results in the  best energy efficiencies and SoC convergence.  We notice extremely high efficiencies because the voltage differences observed in the balancing exercise for this application are on the order of millivolts. From the simulations, we see that the best convergence of SoC  for a full size pack is achieved for a capacitor value of $50F$, resistor value of $0.5 m \Omega$ and a switching time of $0.5 \tau$.  We also note that the initial SoC imbalance between the batteries also determines the time it takes for the charge shuttling to be complete. 

\begin{figure}[htp]
  
  \centering
    \includegraphics[width=1.5in]{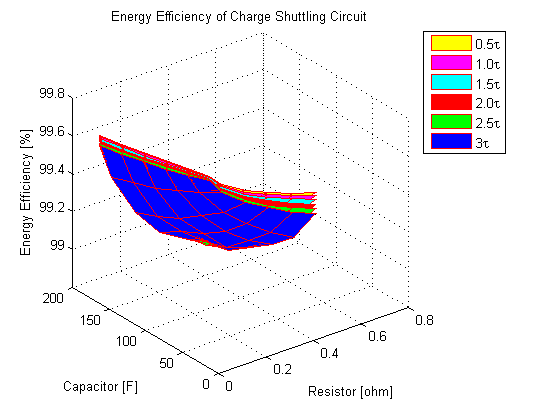}
\caption{Energy efficiency.}
\end{figure}

\begin{figure}[htp]
  
  \centering
    \includegraphics[width=1.5in]{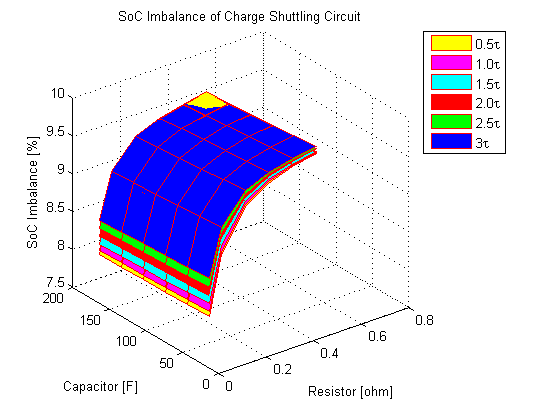}
\caption{SoC imbalance.}
\end{figure}

\section{Simulation of A $3P4S$ Pack}
We now present the simulation results of a $3P4S$ pack in which all the cells are relaxed and at an equilibrium state with 60$\%$ SoC, corresponding to an initial voltage of 3.3 volts.  The individual cells are monitored to determine which cells are unbalanced. In addition, at each time step, the cells having the maximum and minimum SoC are also identified. The cells with the largest imbalance are then isolated and connected to the balancing circuit. The capacitor then accepts charge from the battery with the highest voltage and shuttles it to the battery with the lowest voltage. Charge shuttling is a function of the capacitor value, resistor value, switching frequency, and voltage difference between the unbalanced cells. The highest and lowest cells are re-identified every time step to ensure that the battery cells with the largest imbalance are the ones that are isolated for balancing. The balancing action continues until the maximum SoC deviation is less than 2$\%$.

To observe the cell balancing action, we set the model to simulate a rest time of 12 hours during which the balancing circuit would operate.  The balancing circuit elements were set with a capacitor value of 50F, resistor value of 0.5m$\Omega$ and a switching time factor of 0.5. The simulation time step was set to 0.1. The first half hour of simulation shows the individual battery SoC’s behavior over the active portion of the input current profile. Once the batteries are in a relaxed state at the end of the current profile, we notice a SoC imbalance of a little over 4$\%$ as shown in Fig. 8.
\begin{figure}[htp]
  
  \centering
    \includegraphics[width=2in]{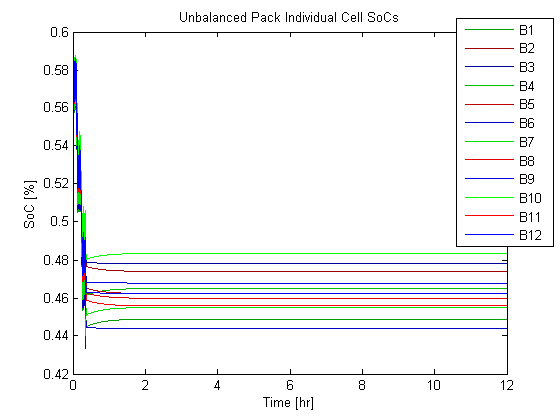}
\caption{SoC of unbalanced $3P4S$ Pack.}
\end{figure}
Fig. 9 shows the individual battery voltages. We see a drift in the voltage as well as the SoC just after one cycle. If left unchecked, this condition would continue to grow leading to degradation of the cell and affect the life and performance of the battery pack. 

\begin{figure}[htp]
  
  \centering
    \includegraphics[width=2in]{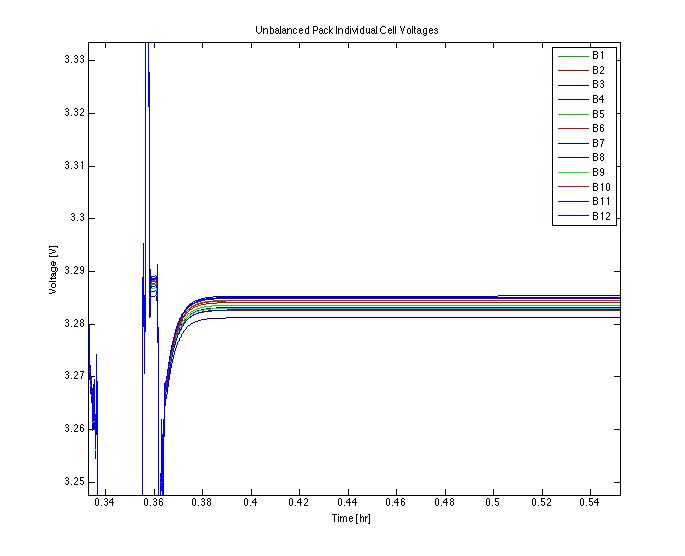}
\caption{Individual cell voltages unbalanced $3P4S$ pack - expanded.}
\end{figure}
Fig. 10 clearly shows a time instant at which the balancing circuit is switching intelligently between multiple high and low voltage batteries. At 2.365 hours, battery $B5$ is connected to battery $B1$. At 2.37 hours, battery $B10$ is connected to $B6$.

\begin{figure}[htp]
  
  \centering
    \includegraphics[width=2in]{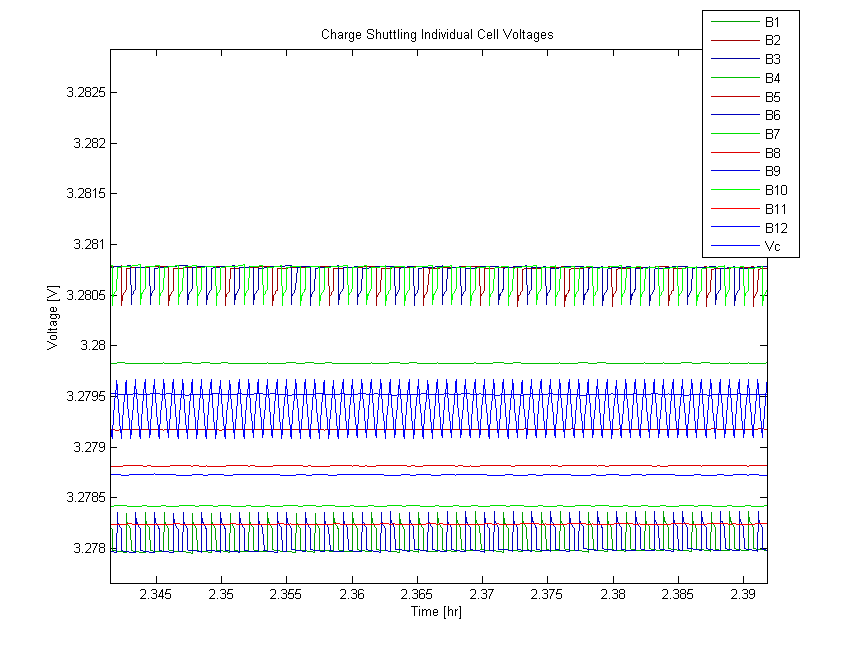}
\caption{Switching action in a $3P4S$ pack.}
\end{figure}
Fig. 11 shows the complete SoC for the pack after balancing. We see that the all the cells are  brought to within a 2$\%$ SoC in 5.1 hours. We note that the SoC at the end of balancing action is lower than the  initial SoC. This is because the we use a charge depleting current profile. 
\begin{figure}[htp]
  
  \centering
    \includegraphics[width=2in]{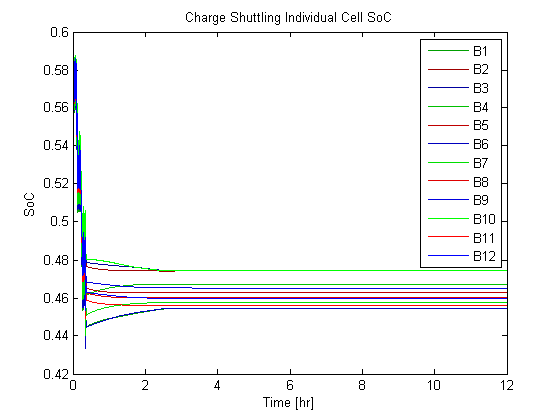}
\caption{SoC of balanced $3P4S$ Pack.}
\end{figure}

\begin{table}[htp]
\centering
\caption{$3P \times 4S$ comparison of battery voltages.}
\begin{tabular}{|c|c|c|c|c|}

  \hline
   Battery & Initial voltage & Final voltage & Delta (mV) & $\%$ difference\\
   \hline
   $B1$ & 3.277 & 3.279 & 2 & 0.0610\\
 $B2$ & 3.280 & 3.279 & 1 &0.0305\\
$B3$ & 3.281 & 3.279 & 2 &0.0610\\
$B4$ & 3.279 & 3.280 & 1 &0.0305\\
$B5$ & 3.279 & 3.280 &1& 0.0305\\
$B6$ & 3.280 & 3.279 & 1 &0.0305\\
$B7$ & 3.278 & 3.278 & 0 &0.0000\\
$B8$ & 3.278 & 3.278 & 0 &0.0000\\
$B9$ & 3.276 & 3.279 & 3 &0.0915\\
$B10$ & 3.282 & 3.280 & 2 &0.0610\\
$B11$ & 3.279 & 3.278 & 1 &0.0305\\
$B12$ & 3.280 & 3.277 & 3 &0.0915\\

   \hline
   
\end{tabular}

\end{table}
Table 4 captures the initial battery voltages at the end of the current profile, the final voltages after balancing action and the percentage difference. We notice a good convergence of the battery voltages at the end of charge shuttling. Notice that the initial voltage imbalance of $5mV$ is reduced to $3mV$ at the end of balancing.

Next, we focus on a pack which has four randomly selected batteries two  of which start  with initial SoC of 64$\% $ and 60$\% $ and  two other batteries with initial SoC of 58$\% $ and 56$\% $. The rest of the batteries in the pack are set to the same initial SoC of 60$\%$. Thus, we have two extremely imbalanced cells, one at overvoltage and one at undervoltage condition, along with two other cells that are mildly imbalanced, again one slightly overvoltaged and the other slightly undervoltaged. The goal of this simulation is to show that the balancing action can be achieved with extremely imbalanced batteries as well.

Fig. 12 shows the individual SoC curves for the balanced pack. Since the balancing continues until the SoC deviation is less than 2$\%$, this is a good metric to observe. We note that within the first few hours, the two most extremely imbalanced cells undergo the most balancing action. It can also be seen that the other two imbalanced cells also start to become balanced. Further analysis of the Fig. 12  shows that at 3 hours the two overvoltaged cells start to balance at the same rate and the same thing occurs for the two undervoltaged cells. The goal of completely balancing the cells to within 2$\%$ SoC  is achieved in 10.4 hours.
\begin{figure}[htp]
  
  \centering
    \includegraphics[width=2in]{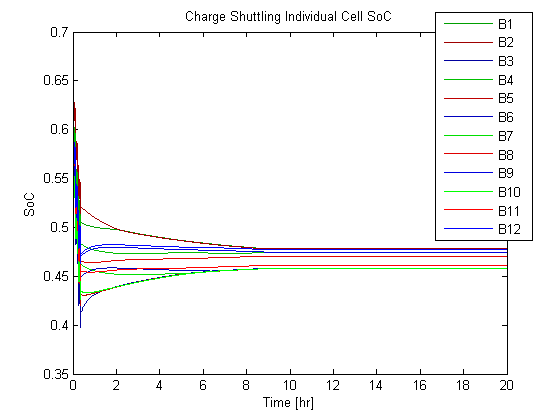}
\caption{SoC of balanced $3P4S$ pack.}
\end{figure}
The final voltage deviation is $3mV$, a reduction of $7mV$ from the initial unbalanced delta of $10mV$.

\section{Conclusion}

Cell equalization of all the cell in a Li-ion parallel pack is fundamental for prolonging the life and the performance of the battery. The procedure mentioned in this work is very similar to other proposed methods to equalize unbalanced cells using switched capacitor method. The main difference lies in applying the method to battery packs with parallel strings. Much of what we found in the literature is for series string of batteries. Advanced vehicle architectures places an increased demand in electric capacity creating a need for strings of batteries to be placed in parallel.  The proposed system is able to equalize the  cell SOC during bidirectional operations, continuously tracking individual cell SOC.The proposed cell equalization method is an attempt to provide a solution that is efficient and  cost effective for a parallel configured battery pack. The method shows a good performance both in speed and accuracy of the equalization process, starting from highly imbalanced cells and bringing the SOC difference among the cells under the fixed threshold during a single charge or discharge cycle. For a pack with extremely imbalanced cells, we achieved a 7mV reduction in final battery voltages in approximately 10 hours.





%

\end{document}